\documentclass[sigconf]{acmart}

\PassOptionsToPackage{hyphens}{url}\usepackage{hyperref}

\usepackage{opera}
\usepackage{booktabs} 
\usepackage[ruled]{algorithm2e} 
\usepackage{balance}

\renewcommand\footnotetextcopyrightpermission[1]{}

\usepackage{etoolbox}
\patchcmd{\maketitle}{\@copyrightspace}{}{}{}

\begin{document}
\title{Viewpoint | Personal Data and the Internet of Things}
\subtitle{It is time to care about digital provenance.}
	
\author{Thomas Pasquier}
\affiliation{%
	\institution{University of Bristol}
}

\author{David Eyers}
\affiliation{%
	\institution{University of Otago}
}

\author{Jean Bacon}
\affiliation{%
	\institution{University of Cambridge}
}	

\begin{abstract}
The Internet of Things promises a connected environment reacting to and addressing our every need, but based on the assumption that all of our movements and words can be recorded and analysed to achieve this end. Ubiquitous surveillance is also a precondition for most dystopian societies, both real and fictional.
How our personal data is processed and consumed in an ever more connected world must imperatively be made transparent, and more effective technical solutions than those currently on offer, to manage personal data must urgently be investigated.
\end{abstract}
	
\maketitle

\noindgras{}

\subsection*{The need for greater transparency}

We have all read market predictions describing billions of devices and the hundreds of billions of dollars in profit that the Internet of Things (\iot) promises.\footnote{\url{https://goo.gl/udt9vh}}
Security and the challenges it represents~\cite{singh2016twenty} are often highlighted as major issues for \iot, alongside scalability and standardisation.
In 2017, FBI Director James Comey warned, during a senate hearing, of the threat represented by a botnet taking control of devices owned by unsuspecting users.
Such a botnet can seize control of devices ranging from
connected dishwashers,\footnote{\url{https://nvd.nist.gov/vuln/detail/CVE-2017-7240}}
to smart home cameras and connected toys, not only using them as a platform to launch cyber attacks, but also potentially harvesting the data such devices collect.

In addition to concerns about cybersecurity, corporate usage of personal data has seen increased public scrutiny. A recent focus of concern has been 
connected home hubs (\eg Amazon Alexa, Google Home).\footnote{\url{https://www.wired.com/2016/12/alexa-and-google-record-your-voice/}}
Articles on the topic discussed whether conversations were being constantly recorded and if so, where those records went.
Similarly, the University of Rennes faced a public backlash after revealing its plan to deploy smart-beds in its accommodation to detect \emph{``abnormal''} usage patterns.\footnote{\url{https://goo.gl/pzC1Kz}}
A clear question emerges from \iot-related fears, \emph{``how and why is my data being used?''}

As concerns grow, legislators across the world are taking action in order to protect the public.
For example, the recent EU General Data Protection Regulation (GDPR) which took effect in May 2018,\footnote{\url{http://www.privacy-regulation.eu/en/index.htm}} and the forthcoming ePrivacy Regulation\footnote{\url{https://ec.europa.eu/digital-single-market/en/proposal-eprivacy-regulation}} place strong responsibility on data controllers to protect personal data, and to notify users of security breaches.
The EU commission defines a Data Controller as the party that determines the purposes for which, and the means by which, personal data is processed (\emph{why} and \emph{how} the data is processed). 
EU regulations further impose constraints on \emph{where} EU citizens data can be processed and \emph{what} type of data (\ie ``special category'' data falls under more stringent constraints).
The data controller must provide means for end users to determine whether their data is properly handled and means to effect their rights.
Overall, there must be mechanisms to determine \emph{what} data is processed, \emph{how}, \emph{why} and \emph{where}.

Such concerns have drawn researchers to look at means to develop more accountable and transparent systems~\cite{crabtree2018building, pasquier2017data}.
The problem has also been clearly highlighted by the EU Data Protection Working Party:  \emph{``As a result of the need to provide pervasive services in an unobtrusive manner, users might in practice find themselves under third-party monitoring. This may result in situations where the user can lose all control on the dissemination of his/her data, depending on whether or not the collection and processing of this data will be made in a transparent manner or not.''}

Indeed, modern computing systems contain many components that operate as black boxes;
they accept inputs and generate outputs but do not disclose their internal working.
Beyond privacy concerns, this also limits the ability to detect cyber-attacks, or more generally to understand cyber-behaviour.
Because of these concerns DARPA, in the US, launched the \emph{Transparent Computing} project\footnote{\url{https://www.darpa.mil/program/transparent-computing}} to explore means to build more transparent systems through the use of digital provenance with the particular aim of identifying advanced persistent threats.
While DARPA's work is a good start, we believe that there is an urgent need to reach much further.
In the rest of the article, we explore how provenance can be an answer to some \iot concerns and the challenges faced to deploy provenance techniques.

\subsection*{Digital Provenance}

There is a growing clamour for more transparency, but straightforward, widespread technical solutions have yet to emerge.
Typical software log records often prove insufficient to audit complex distributed systems as they fail to capture the complex causality relationships between events.
Digital provenance~\cite{carata2014primer} is an alternative means to record system events. Digital provenance is the record of information flow within a computer system in order to assess the origin of data (\eg its quality or its validity).

The concept first emerged in the database research community as a means to explain the response to a given query~\cite{herschel2017survey}.
Provenance research later expanded to address issues of scientific reproducibility, notably by providing mechanisms to reconstitute computational environments from formal records of scientific computations~\cite{pasquier2017if}.
More recently, provenance has been explored within the cybersecurity community~\cite{pohly2012hi} as a means to explain intrusions~\cite{king2003backtracking} or more recently to detect them~\cite{han2017}.

Provenance records are represented as a directed acyclic graph that shows causality relationships between the states of the objects that compose a complex system.
As a consequence, it is compatible with automated mathematical reasoning.
In such a graph, the vertices represent the state of transient and persistent data items, transformations applied to those states, and persons (legal or natural) responsible for data and transformations (generally referred to as entities, activities and agents respectively).
The edges represent dependencies between these entities.
The analysis of such a graph allows us to understand \emph{where}, \emph{when}, \emph{how}, \emph{by whom} and \emph{why} data has been used~\cite{cheney2009provenance, buneman2001and}.

An outcome of research on provenance in the cybersecurity space is the understanding that the capture mechanism must provide guarantees of completeness (\ie all events in the system can be seen), accuracy (\ie the record is faithful to events) and a well-defined, trusted computing base (\ie the threat model is clearly expressed)~\cite{pasquier2018ccs}.
Otherwise, attacks on the system may be undetected, dissimulated
by the attacker or misattributed.
We argue that in a highly \emph{ad hoc} and interoperable environment with mutually untrusted parties, the provenance used to empower end users with control and understanding over data usage requires similar properties.

\subsection*{Who to trust?}

In the \iot environment the number of involved stakeholders has the potential to explode exponentially.
Traditionally, a company managed its own server infrastructure, maybe with the help of a subcontractor.
The cloud computing paradigm further increased complexity with the involvement of cloud service providers (sometimes stacked, \eg \textsf{Heroku} PaaS on top of the \textsf{Amazon} IaaS cloud service),
third party service providers (\eg \textsf{CloudMQTT}) and other tenants sharing the infrastructure.
The \iot further increases this complexity, with potentially \emph{ad hoc} and unforeseen interactions between devices and services on top of the complex cloud and edge computing infrastructure most \iot services rely on.

One answer to this problem is to build applications in ``silos'' where the involved parties are known in advance, but as a side-effect locking-in devices and services to a single company (\eg the competing smart-home offerings by leading technology companies).
This is far from the \iot vision of a connected environment, but most existing products fall in this category.
There are obviously major business considerations behind this model, and it should be noted that the EU GDPR mandates for some form of interoperability (although it is yet unclear how it should be interpreted~\cite{de2017right}).

An alternative to such \emph{``lock-in''} would be to make devices' consumption of data transparent and accountable.
If data is exchanged across devices, the concerned user should be able to audit its usage.
However, in an environment where arbitrary devices could interact (although it must be remembered that EU GDPR requires explicit and informed user consent), how can trust be established in the audit record?
This requires an in-depth rethinking of how \iot platforms are designed, potentially exploring the security-by-design approach based on hardware roots of trust~\cite{eldefrawy2012smart} to provide trusted digital enclaves in which behaviour can be audited and to encourage some form of \emph{``accountability-by-design''} principles where transparency and the implementation of a trustworthy audit mechanism is a core concern in product design.

Such solutions have been explored in the provenance space,
for example, by leveraging Software Guard Extensions (SGX) properties to provide a strong guarantee of the integrity of the provenance record~\cite{balakrishnan2017non}.
Similarly, remote attestation techniques leveraging Trusted Platform Module (TPM) hardware have been proposed~\cite{bates2015trustworthy} to guarantee the integrity of the capture mechanism.
However, how to provide such guarantees in an \iot environment, where such hardware features may not be available, is a relatively unexplored topic.

\subsection*{Where does the audit live?}

The fully realised \iot vision is of vast distributed and decentralised systems.
If we assume trustworthy provenance capture is achievable, the issue of guaranteeing that the provenance record can be audited remains. 
If you are to audit the processing of personal data, guarantees about the integrity and availability of the provenance record must exist.
If you agreed to share your daily activity for research,
the activities of insurance companies scraping your data for possible health risks
must not be able to masquerade as benign research use, nor should data collection for political purposes be able to pass as harmless entertainment, as in the Cambridge Analytica scandal.\footnote{\url{https://goo.gl/pPXhZ1}}
Similarly, the availability (durability) of the audit record must be guaranteed. There is no point to an audit record if it can simply be deleted.

Further, Moyer \etal evaluated the storage requirements of provenance when used for security purposes in relatively modest distributed systems~\cite{moyer2016high}.
In such a context, several thousands of graph elements can be generated per second and per machine, resulting in a graph containing billions of nodes to represent system execution over several months.
It is unclear how some past research outcomes: \eg
detection of suspicious behaviour~\cite{allen2011provenance}, privacy-aware provenance~\cite{davidson2011provenance} or provenance integrity~\cite{hasan2009case}; scale to very large graphs as such concerns were not evaluated.
Similarly, while blockchain is heralded~\cite{liang2017provchain} as an integrity-preserving means to store provenance, it is unclear how well it could expand to such scale.
Several options have been explored to reduce graph size, such as identifying and tracking only sensitive data objects~\cite{bates2015take} or performing property-preserving graph compression~\cite{hossain18dependence} however none has yet adequately addressed the scalability challenge.

\subsection*{How to communicate information?}

Means must be developed to communicate about data usage, but also about the risks of inference from the data. 
Not only must the nature of the data be considered, but also other properties such as the frequency of capture~\cite{amar2018information}.
For example, a 100\,Hz smart meter reading can in some cases indicate what television channel is currently being watched; even a daily average reading could inform about occupancy.
Here, it is important to be able to explore and represent the outcome of complex computational workflow~\cite{acar2010graph}.

Provenance visualisation has been an active research topic for over a decade, yet no fully satisfactory solution has been proposed.
The simplest possible visualisation is to render the graph, however beyond trivially simple graphs such a representation is too complex and dense to be easily understood, even by experts.
We go further and suggest that educational background, socio-economic environment and culture may play a part in how interpretable such information is.

In order to make the accountability and transparency of \iot platforms effective, a better communication medium must be provided.
An approach often taken is to analyse motifs in the graph to extract high-level abstractions (\eg Missier \etal~\cite{missier2014provabs}), meaningful to the average end-user.
In recent work~\cite{schreiber2017tracing}, it was proposed to represent such a high-level abstractions as a comic strip.

\subsection*{We need to care about digital provenance}

Building transparent and auditable systems may be one of the greatest software engineering challenges of the coming decade.
As a consequence, digital provenance and its application to cybersecurity and the management of personal data has become a hot research topic.
We have highlighted key active areas of research and their associated challenges.
It is fundamental for industry practitioners to understand the threat posed by the black-box nature of the \iot, the potential solutions, and the challenges to a practical deployment of those solutions.
Accountability-by-design must become a core objective of \iot platforms.

\vspace{0.5cm}
\noindgras{Thomas Pasquier} (\url{http://tjmp.org}) is a Lecturer (Assistant Professor) at the University of Bristol in the Cyber Security Group, and a visitor at the University of Cambridge in the Department of Computer Science and Technology.

\noindgras{David Eyers} (\url{https://www.cs.otago.ac.nz/staff/David_Eyers}) is an Associate Professor in the Department of Computer Science at the University of Otago.

\noindgras{Jean Bacon} (\url{http://www.cl.cam.ac.uk/~jmb25/}) is Professor Emerita of Distributed Systems at the University of Cambridge.

\balance
\bibliographystyle{ACM-Reference-Format}
\bibliography{biblio}

\end{document}